\documentclass[12pt]{article}
\usepackage{epsf}

\begin{document}
\begin{titlepage}
\def\thepage {}    % Kill page numbering

\title{Terrestrial Neutrino Oscillations Illustrated\thanks{Research supported
in
part by the
National Science Foundation
under grant number NSF-PHY/98-02709.}\,\,\thanks{HUTP-99/A038}
}

\author{
Howard Georgi\thanks{\tt georgi@physics.harvard.edu}\,\,
and Sheldon L. Glashow\thanks{\tt glashow@boyle.harvard.edu}\\
Lyman Laboratory of Physics\\
Harvard University\\
Cambridge, MA 02138}

\date{July, 1999}

\maketitle

\vspace{24pt}

\begin{abstract}
\pagestyle{empty}

Observations of atmospheric neutrinos offer compelling evidence that
neutrinos have mass and do oscillate. Preliminary data are compatible with
maximal $\nu_\mu$--$\nu_\tau$ mixing, but not with pure
$\nu_\mu$--$\nu_e$ mixing. In
a general three-family scenario with just one relevant 
squared-mass difference, atmospheric neutrino
oscillations involve two
mixing angles. The special cases mentioned above
are not favored by
convincing theoretical arguments. As more precise data 
are accumulated, both at 
Superkamiokande and
at proposed or ongoing long-baseline experiments, it will become 
feasible and desirable
to measure both angles. To this end, we offer a brief portfolio of
illustrations
from which the qualitative effects of the two mixing angles 
on various observable quantities can be discerned.

\end{abstract}
\end{titlepage}

%%%%%%%%%%%%%%%%%%%%%%%%%%%%%%%%
\setcounter{section}{0}
\setcounter{equation}{0}

Observations of atmospheric and solar neutrinos suggest that
neutrinos have mass and are subject to flavor oscillations. These 
oscillations may be described in terms of three chiral neutrinos with
squared-mass differences $\Delta_{ij}\equiv |m_i^2-m_j^2|$ satisfying:
\begin{equation}
\label{edelta}{ \Delta_{13}\simeq \Delta_{23}\simeq 10^{-3}\;{\rm eV^2}\,,
\quad\ {\rm and}\quad\  \Delta_{12}\le 10^{-5}\; {\rm eV^2}\,.}
\end{equation}

The smaller difference $\Delta_{12}$ is relevant to solar neutrino
oscillations, but it hardly affects oscillations of
atmospheric neutrinos or of neutrinos to be studied at long-baseline
experiments. These `terrestrial oscillations' involve 
the larger squared-mass difference. 
They depend on two mixing angles, which we denote by $\theta_1$ and
$\theta_2$, parametrizing the decomposition of the mass eigenstate $\nu_3$
into lepton flavor eigenstates:
\begin{equation}
\nu_3=s_2\,\nu_e+s_1c_2\,\nu_{\mu}+c_1c_2\,\nu_{\tau}
%\label{}
\end{equation}
where $s_i$ and $c_i$ stand for sines and cosines of $\theta_i$.
The relevant
flavor-transition probabilities are:
\begin{equation}\label{etp}\begin{array}{r@{\;}l}
 P(\nu_\mu \rightarrow\nu_\tau)\simeq & 4B(E)\,
s_1^2 c_1^2 c_2^4\,,\qquad
 P(\nu_e \leftrightarrow\nu_\mu)\simeq 4B(E)\,
s_1^2 s_2^2 c_2^2\,,\\
& P(\nu_e \rightarrow\nu_\tau)\simeq 4B(E)\,
c_1^2s_2^2c_2^2 \,,\\
\end{array}\end{equation}
where
\begin{equation}
\label{eB}{B\equiv \sin^2{(\Delta_{13} R/4E)}\,,}
\end{equation}
with $E$ the neutrino energy and $R$ its flight length.
These results are familiar~\cite{rfam} and have been used to perform
extensive analyses of available data~\cite{rfam,rmany}.\footnote{The angles
$\theta_1$ and $\theta_2$ correspond to $\theta_{23}$ and $\theta_{13}$
respectively in reference \cite{rfam}.} Our very modest purpose in this
note is simply to exhibit how various observables depend on the two
mixing angles. 

We begin by considering atmospheric neutrino oscillations. 
Let $N_\mu$ and $N_e$ be the fluxes of $e$-like and $\mu$-like events
that
would be seen at a given site and direction
were there no oscillations. The observed (primed) fluxes will be:
\begin{equation}\label{eaflux}\begin{array}{r@{\;}l}
N_\mu'&= \big(1-4B\,s_1^2c_2^2(1-s_1^2c_2^2)\big)\,N_\mu +
4B\,s_1^2s_2^2c_2^2\,N_e\,,\\ 
N_e' &= (1-4B\,s_2^2c_2^2)\,N_e + 4B\,s_1^2s_2^2c_2^2\,N_\mu\,.\\
\end{array}\end{equation}
To develop a feel for the import of these equations, 
we examine them in following simple limit:
 We use the approximation $N_\mu=2N_e$ and replace
$B$ by its time and energy averaged value of 1/2. With these
substitutions, the often-considered ratio of ratios becomes:
\begin{equation}
\label{err}{R\equiv(N_\mu'/N_e')/(N_\mu/N_e)=
{1-2s_1^2c_2^2+2s_1^4c_2^4+s_1^2s_2^2c_2^2 \over 1 +4s_1^2s_2^2c_2^2
-2s_2^2c_2^2}\,, }
\end{equation}
a quantity that can vary within the interval $2\ge R\ge 0.5$. The
observations are compatible with a value of $R$ near its lower bound.
Indeed, that bound may be achieved at just two points: with 
maximal $\nu_\mu$--$\nu_\tau$ mixing ($s_1^2=1/2, \ s_2^2=0$), or with
maximal $\nu_\mu$--$\nu_e$ mixing ($s_1^2=1, \ s_2^2=1/2$), the
latter possibility being strongly disfavored. To exhibit the dependence of
this ratio on other values of the mixing angles, we show contour plots for
$R$ in figure~1.

The acceptances and biases of $e$-like and $\mu$-like events at
SuperKamiokande may not be the same. Thus,
when adequate data is available,
it may be desirable to 
study their angular distributions separately. To this end, we
present figures~2 and 3. We note in passing that SuperKamiokande 
data~\cite{rSK} may suggest a small
up/down asymmetry of $e$-like events, and correspondingly, a non-zero value
of $s_2$. The observables displayed in figures~1--3 are independent of the
overall flux of atmospheric neutrinos, which is presently rather uncertain.
Figure~4 shows the effect of oscillations on this quantity. If the flux
uncertainty can be substantially reduced, the event rate may provide a
useful constraint on the mixing angles. 

At least
two long baseline experiments will shed further light on the 
neutrino squared-mass difference and the two mixing angles:
the ongoing K2K experiment and the approved Minos experiment. Figure~5 shows
how the ratio of 
$e$-like events ($M_e$) to all observed events ($M_\mu+M_e$)
depends on $\theta_i$.
(Here we assume that the beam is pure $\nu_\mu$. In fact, there will be a
1--2\%\ admixture of $\nu_e$, but this will be measured at the near
detector.) An observed excess of $e$-like events would prove that 
$s_2\ne 0$ and thus, that all three flavors participate in the oscillations.
Figure~6 shows the ratio of the actual event rate ($M_\mu+M_e$)
 to what it would be in the absence of oscillations---a quantity that would
be useful only if the neutrino beam is precisely directed
toward its distant target. 
Figures~5 and 6 must
be taken with a grain of salt: the replacement of $B(E)$ by its $R/E$
average is not justified for either experiment. Our illustrations are
intended solely as guides to the mind.

\newpage
{\begin{figure}[p]
{\centerline{\epsfxsize=16cm \epsfbox[153 369 459 657]{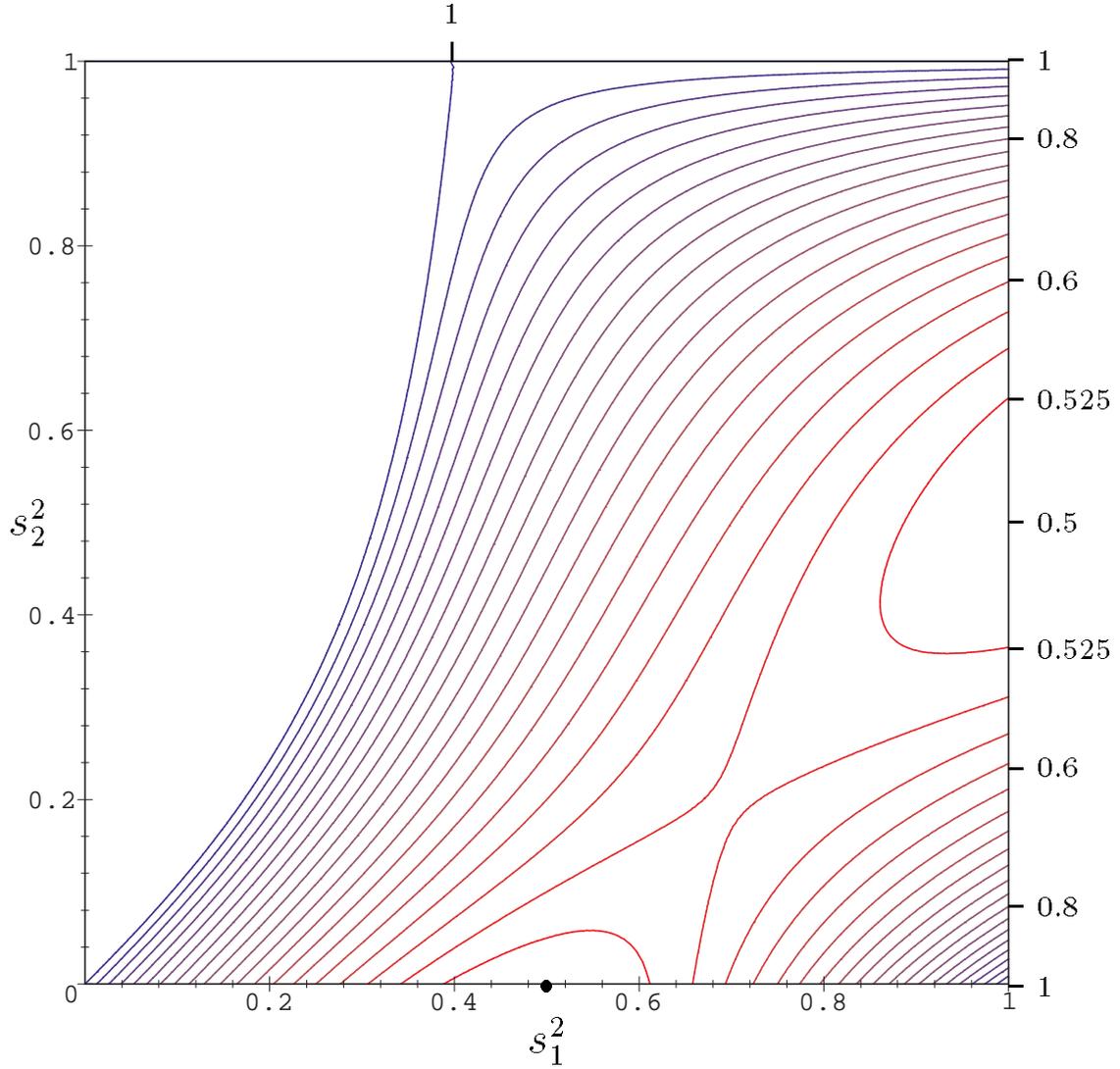}}}
\caption{\sf\label{fig-1} The ratio of ratios of atmospheric neutrino fluxes, 
$R\equiv (N_e'N_\mu/N_\mu'N_e)$ (shown from $1/2$ to $1$ with contour spacing
of 0.025).
Note the shallow valley connecting the two points at
which $R$ assumes its minimum value. The bullet indicates
maximal $\nu_\mu$--$\nu_\tau$ oscillations.
 Figures 1--4 correspond to the assignments
 $B=0.5$ and $N_\mu=2N_e$.}\end{figure}}
{\begin{figure}[p]
{\centerline{\epsfxsize=16cm \epsfbox[153 369 459 657]{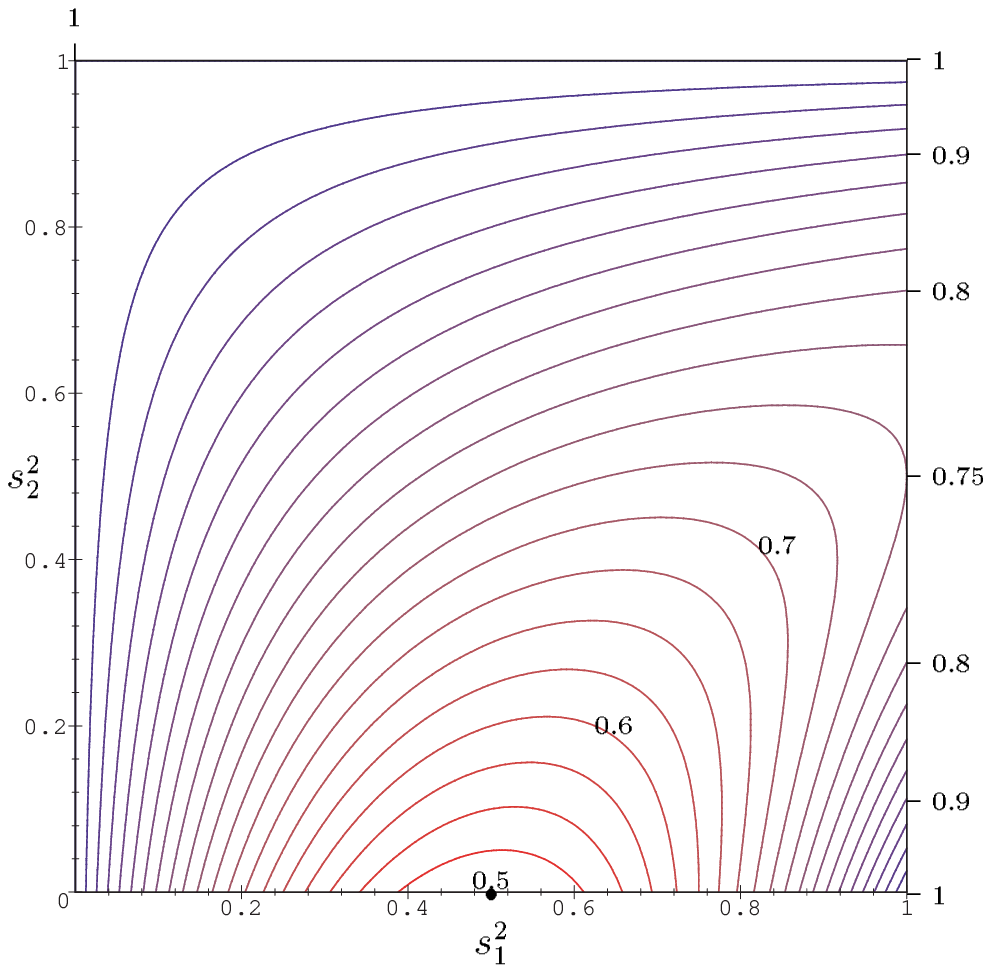}}}
\caption{\sf\label{fig-2} The ratio $N_\mu'/N_\mu$ (it runs from $1/2$ to $1$
with contour spacing 0.025).
This flux-independent 
quantity may be determined from the angular distribution of $\mu$-type
atmospheric neutrino events.}\end{figure}}
{\begin{figure}[p]
{\centerline{\epsfxsize=16cm \epsfbox[153 369 459 657]{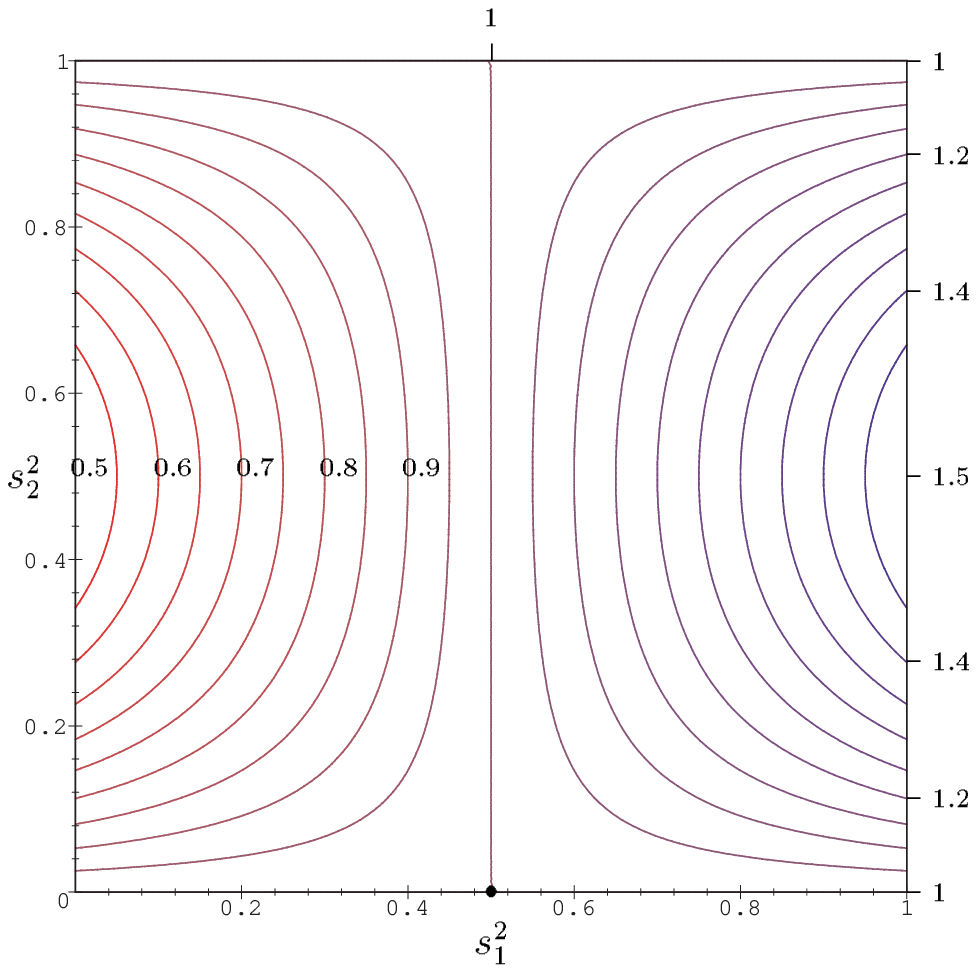}}}
\caption{\sf\label{fig-3} The ratio $N_e'/N_e$ (it runs from $1/2$ to $3/2$
with contour spacing 0.05). This flux-independent 
quantity may be determined from the angular distribution of $e$-type
atmospheric neutrino events.}\end{figure}}
{\begin{figure}[p]
{\centerline{\epsfxsize=16cm \epsfbox[153 369 459 657]{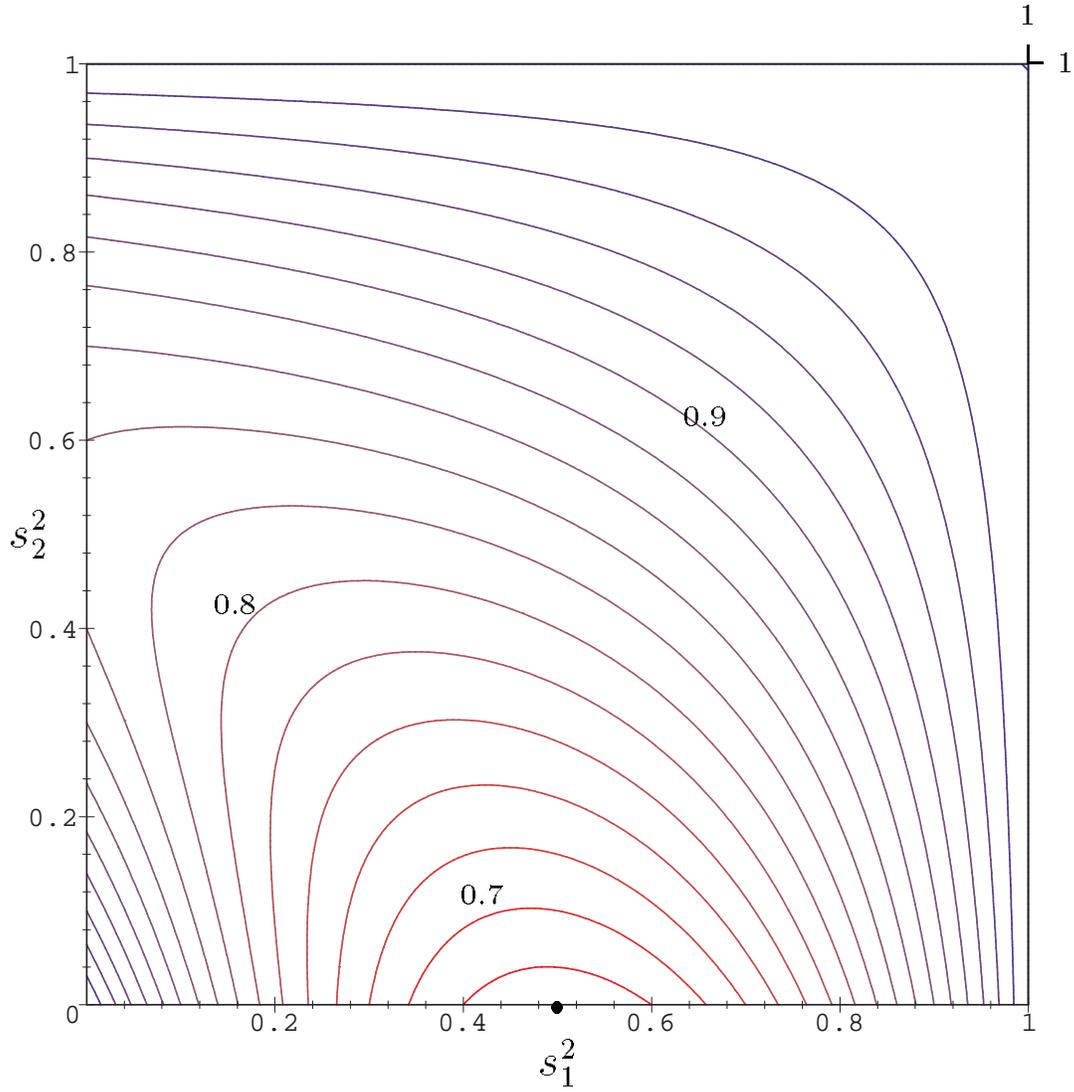}}}
\caption{\sf\label{fig-4} The overall atmospheric neutrino flux,
$(N_\mu'+N_e')/(N_\mu+N_e)$
normalized to its no-oscillation expectation
(it runs from $2/3$ to $1$ with contour spacing 0.02). Current flux
uncertainties prevent 
its precise determination.}\end{figure}}
{\begin{figure}[p]
{\centerline{\epsfxsize=16cm \epsfbox[153 369 459 657]{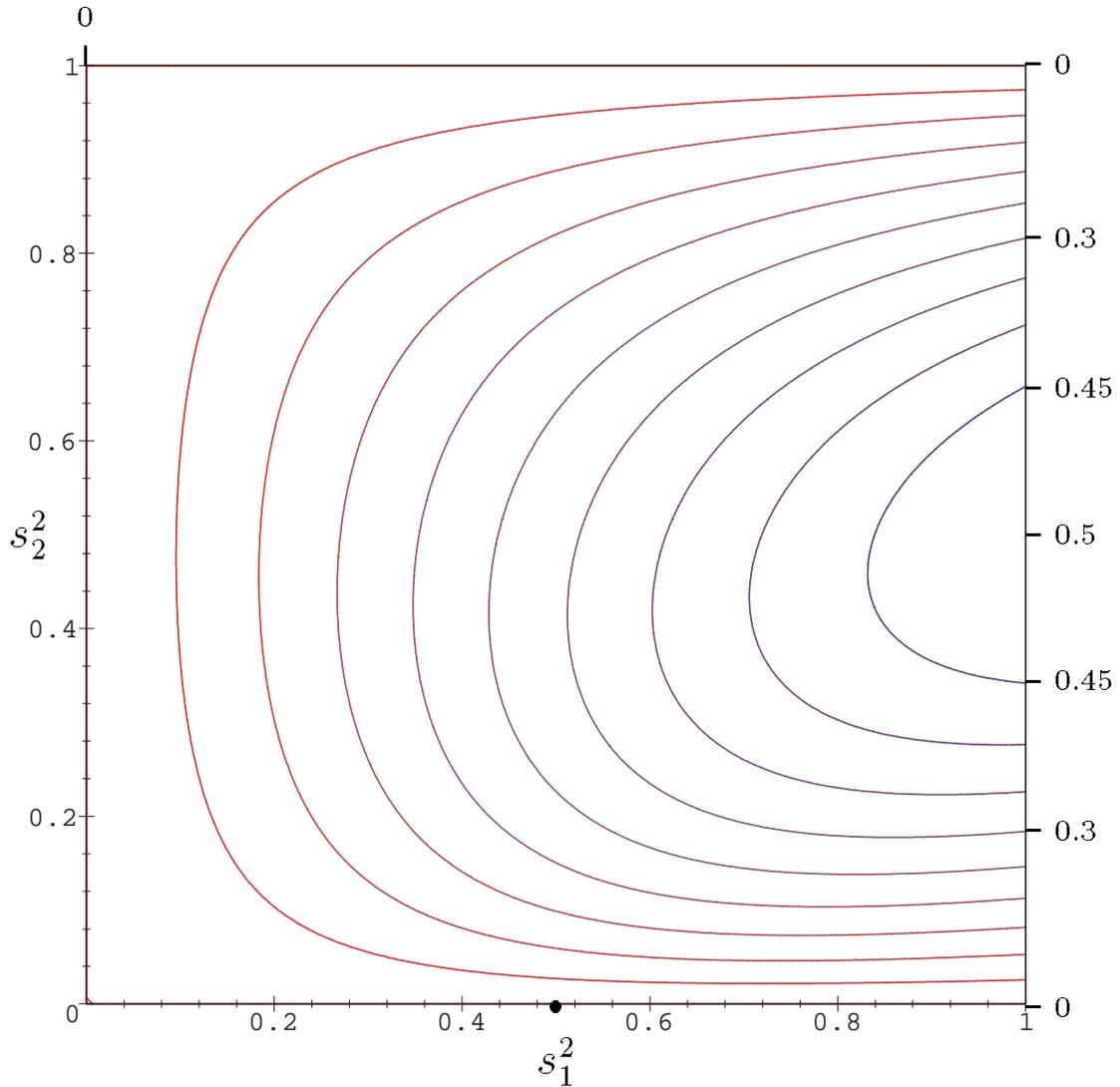}}}
\caption{\sf\label{fig-5} The flux-independent
ratio $M_e/(M_\mu+M_e)$ of $e$-like events to all events
in an imagined very long baseline experiment, where $B(E) $ may be replaced
by its average value of 1/2 (the ratio runs from $0$ to $1/2$ with contour
spacing 0.05).
In real experiments such as K2K and Minos,
this replacement may not be implemented. }\end{figure}}
{\begin{figure}[p]
{\centerline{\epsfxsize=16cm \epsfbox[153 369 459 657]{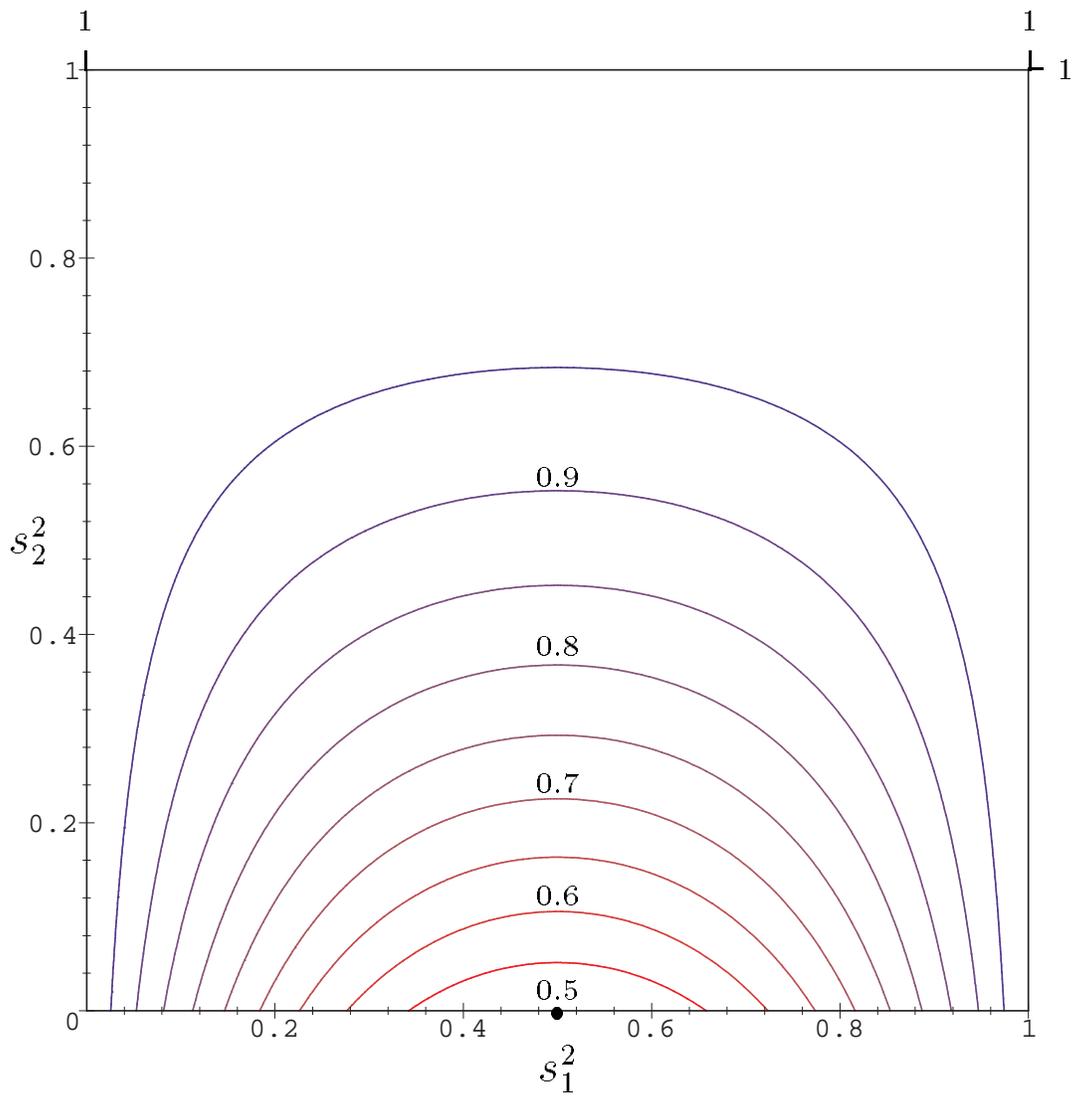}}}
\caption{\sf\label{fig-6} The ratio of the observed event rate $M_\mu+M_e$ to
its value with
no oscillations, again with $B=1/2$ (the ratio runs from $1/2$ to $1$ with
contour spacing 0.05).}\end{figure}}

\end{document}